\documentclass{article}

\def\xxinput#1{\input#1}

\xxinput{vsolj02.sty}

\usepackage[dvipdfmx]{graphicx}
\usepackage[comma,colon]{natbib}

\usepackage[OT2,T1]{fontenc}

\def\cite{\citealt}
\setcitestyle{aysep={}}

\def\commenta{$^*$}
\def\commentb{$^\dagger$}

\newcounter{author}
\def\altaffilmark#1{$^{#1}$}
\def\altaffiltext#1{$^{#1}$\,}

\def\authorcount#1#2{{\refstepcounter{author}\label{#1}
                     \altaffiltext{\ref{#1}}{#2}}}

\begin{document}

\begin{center}

\title{ASASSN-22ak: \textbf{\textit{La Belle au bois dormant}} in a hydrogen-depleted dwarf nova?}

\author{
        Taichi~Kato\altaffilmark{\ref{affil:Kyoto}},
        Franz-Josef~Hambsch\altaffilmark{\ref{affil:GEOS}}$^,$\altaffilmark{\ref{affil:BAV}}$^,$\altaffilmark{\ref{affil:Hambsch}}
        Berto~Monard,\altaffilmark{\ref{affil:Monard}}$^,$\altaffilmark{\ref{affil:Monard2}}
        Rod~Stubbings\altaffilmark{\ref{affil:Stubbings}}
}

\authorcount{affil:Kyoto}{
     Department of Astronomy, Kyoto University, Sakyo-ku,
     Kyoto 606-8502, Japan \\
     \textit{tkato@kusastro.kyoto-u.ac.jp}
}

\authorcount{affil:GEOS}{
     Groupe Europ\'een d'Observations Stellaires (GEOS),
     23 Parc de Levesville, 28300 Bailleau l'Ev\^eque, France \\
     \textit{hambsch@telenet.be}
}

\authorcount{affil:BAV}{
     Bundesdeutsche Arbeitsgemeinschaft f\"ur Ver\"anderliche Sterne
     (BAV), Munsterdamm 90, 12169 Berlin, Germany}

\authorcount{affil:Hambsch}{
     Vereniging Voor Sterrenkunde (VVS), Oostmeers 122 C,
     8000 Brugge, Belgium}

\authorcount{affil:Monard}{
     Bronberg Observatory, Center for Backyard Astrophysics Pretoria,
     PO Box 11426, Tiegerpoort 0056, South Africa \\
     \textit{astroberto13m@gmail.com}}

\authorcount{affil:Monard2}{
     Kleinkaroo Observatory, Center for Backyard Astrophysics Kleinkaroo,
     Sint Helena 1B, PO Box 281, Calitzdorp 6660, South Africa}

\authorcount{affil:Stubbings}{
     Tetoora Observatory, 2643 Warragul-Korumburra Road, Tetoora Road,
     Victoria 3821, Australia \\
     \textit{stubbo@dcsi.net.au}
}

\end{center}

\begin{abstract}
\xxinput{abst.inc}
\end{abstract}

\section{Introduction}

   In the famous fairy tale \textit{La belle au bois dormant}
(the Beauty in the Sleeping Forest or the Sleeping Beauty),
a princess was cursed by an evil fairy to sleep for a hundred years
before being awakened by a prince \citep{per1697sleepingbeauty}.
This tale produced one of the world most famous ballets composed
by Pyotr Tchaikovsky \citep{tch1889sleepingbeauty}\footnote{
   The reference refers to the earliest publication of this work
   in the form of a score of Aleksandr Ziloti's arrangement for
   solo piano according to Tchaikovsky's letter
($<$https://en.tchaikovsky-research.net/pages/The\_Sleeping\_Beauty$>$).
   The premiere at the Mariinsky Theatre was performed in 1890.
}.  The similar things appear to have happened in the world
of dwarf novae.  The giant outburst and subsequent superoutbursts
in V3101 Cyg = TCP J21040470$+$4631129
\citep{tam20v3101cyg,ham21DNrebv3101cyg} could be a signature
of long ``dormant'' phase before the initial outburst.
MASTER OT J030227.28$+$191754.5 \citep{tam23j0302,kim23j0302}
might be another such example.  Here, we report on an instance
of ASASSN-22ak, which may be the first similar case in
a cataclysmic variable (CV) with an evolved core in the secondary.

\section{ASASSN-22ak}

   ASASSN-22ak was discovered as a dwarf nova by
the All-Sky Automated Survey for
Supernovae (ASAS-SN: \cite{ASASSN}) at $g$=15.0 on 2022 January 7.\footnote{
   $<$https://www.astronomy.ohio-state.edu/asassn/transients.html$>$.
}  The object further brightened and reached the peak of
$g$=13.2 on 2022 January 8.  The object apparently faded
rapidly after this (there was a 6-d gap in observation in
ASAS-SN).  When the object was observed again on 2022 January 16
by Gaia (=Gaia22afw)\footnote{
   $<$http://gsaweb.ast.cam.ac.uk/alerts/alert/Gaia22afw/$>$.
}, the object faded to $G$=15.16.
This outburst was announced in VSNET \citep{VSNET} by
Denis Denisenko (vsnet-alert 26518)\footnote{
   $<$http://ooruri.kusastro.kyoto-u.ac.jp/mailarchive/vsnet-alert/26518$>$.
}. According to this, this outburst was also detected by
MASTER-OAFA \citep{lip10MASTER} at 13.8 mag on 2022 January 9.
The object underwent another outburst
at 15.4 mag on 2022 July 20 detected by one of the authors (RS)
(vsnet-alert 26875)\footnote{
   $<$http://ooruri.kusastro.kyoto-u.ac.jp/mailarchive/vsnet-alert/26875$>$.
} and 16.2 mag on 2022 December 18 (by RS, vsnet-alert 27223)\footnote{
   $<$http://ooruri.kusastro.kyoto-u.ac.jp/mailarchive/vsnet-alert/27223$>$.
}.  After these two outbursts, the unusual light curve of
this object received attention (vsnet-alert 27224).\footnote{
   $<$http://ooruri.kusastro.kyoto-u.ac.jp/mailarchive/vsnet-alert/27224$>$.
}  The ASAS-SN light curve suggested that all outbursts
were superoutbursts.  Although the similarity to V3101 Cyg and
the possibility of an AM CVn star, as judged from the short
recurrence time of long outbursts, were discussed,
the nature of the object remained elusive.
One of the authors (BM) obtained a single-night run during
the 2022 January outburst and a possible period of 0.044~d
was suggested (vsnet-alert 27225).\footnote{
   $<$http://ooruri.kusastro.kyoto-u.ac.jp/mailarchive/vsnet-alert/27225$>$.
}  This period, however, did not comfortably fit what is
expected for a WZ Sge star and the reality of the period
remained to be confirmed.  During the 2022 December outburst,
one of the authors (FJH) obtained time-resolved photometry,
which also suggested a period of 0.0412~d (vsnet-alert 27243).\footnote{
   $<$http://ooruri.kusastro.kyoto-u.ac.jp/mailarchive/vsnet-alert/27243$>$.
}  This suggestion of a period, however, remained unconfirmed
since the object faded soon after these observations
and the amplitudes of the variations were small.
The sudden fading of 1.8~mag (corresponding to more than
2.0 mag d$^{-1}$) on 2022 December 29 was sufficient to
convince us that the 0.0412~d, but not its double, is
the true period (vsnet-alert 27258).\footnote{
   $<$http://ooruri.kusastro.kyoto-u.ac.jp/mailarchive/vsnet-alert/27258$>$.
}

   These outbursts, however, left us important lessons
and we started observations following the detection of
another outburst at 15.2 mag on 2023 April 29 by RS.
FJH obtained time-resolved photometry.  The sampling rate,
however, was initially insufficient to detect a period.
After increasing the sampling rate on 2023 May 13,
the detected period was confirmed to be the same as in
the previous outbursts.
The log of observations is summarized in table \ref{tab:obslog}.

\xxinput{obslog.inc}
\addtocounter{table}{-1}
\xxinput{obslog2.inc}

\section{Long-term behavior}

\begin{figure*}
\begin{center}
\includegraphics[width=16cm]{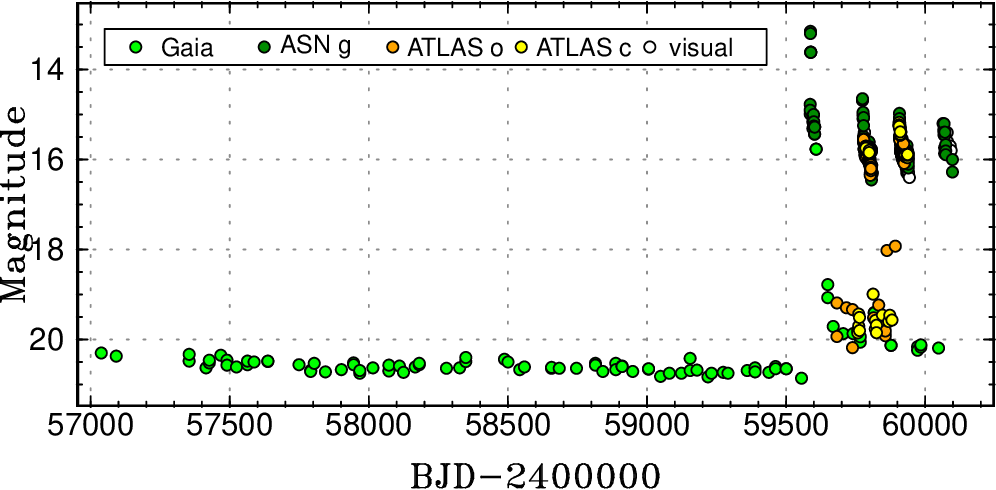}
\caption{
   Long-term light curve of ASASSN-22ak.
}
\label{fig:lclong}
\end{center}
\end{figure*}

\begin{figure*}
\begin{center}
\includegraphics[width=16cm]{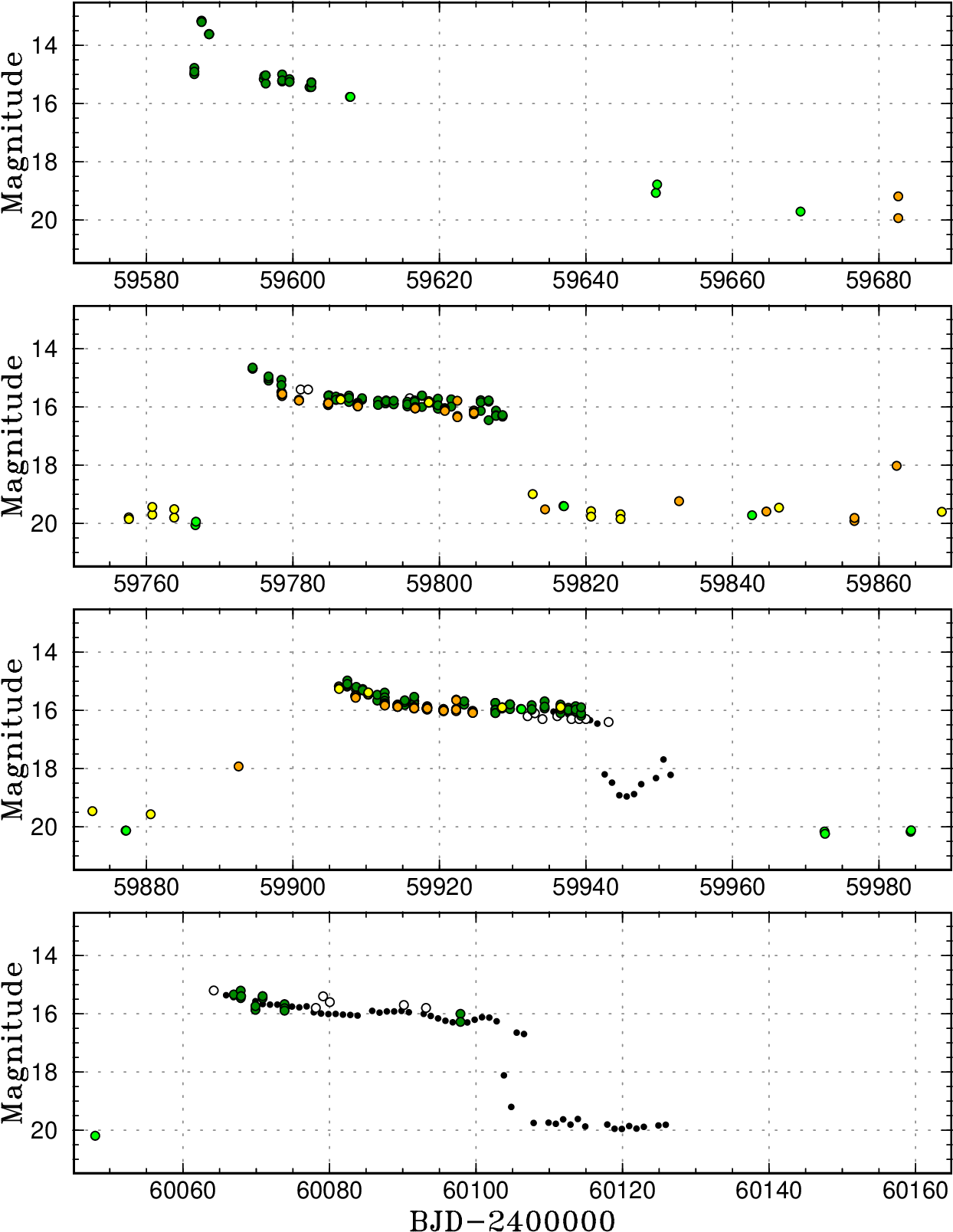}
\caption{
   Light curves the outbursts of ASASSN-22ak.  The peaks of
   the outbursts are not aligned to show the existing data
   between the outbursts better.  The symbols are the same as in
   figure \ref{fig:lclong}.  The small dots represent nightly
   averaged unfiltered CCD magnitudes by FJH.
   Near the termination of the third and fourth outbursts,
   there were a short (0.5--2~d) dip and rebrightening.
   ASAS-SN did not detect the object 1-d before the initial
   detection (upper limit 17.0 mag).
}
\label{fig:lcout}
\end{center}
\end{figure*}

   The long-term light curve of ASASSN-22ak using the survey
data and visual observations by RS is shown in
figure \ref{fig:lclong}.  During Gaia observations between
2015 and 2021, the object very slowly faded.  This trend
was different from V3101 Cyg before the first outburst
\citep{tam20v3101cyg}.
The four outbursts starting from 2022 January are seen
in the right part of the figure.  The quiescent brightness
between these outbursts were brighter than Gaia observations
before the first outburst.  The enlarged light curves of
these outbursts are given in figure \ref{fig:lcout}.
Near the termination of the third and fourth outbursts,
there were a short (less than 1~d in the third and 2~d
in the fourth) dip and rebrightening.  The presence of such
a short dip indicates that the long outbursts were indeed
superoutbursts of a system with a short orbital period,
not long outbursts seen in SS Cyg stars.
We should note that the post-outburst observations after
the 2022 December outburst (third panel in figure \ref{fig:lcout})
were biased brighter since aperture photometry could
measure the object only on limited number of frames.
The true magnitudes should be fainter (see the fourth panel
in figure \ref{fig:lcout}, which were observed under more
ideal conditions).

\section{Superhumps}

\begin{figure*}
\begin{center}
\includegraphics[width=14cm]{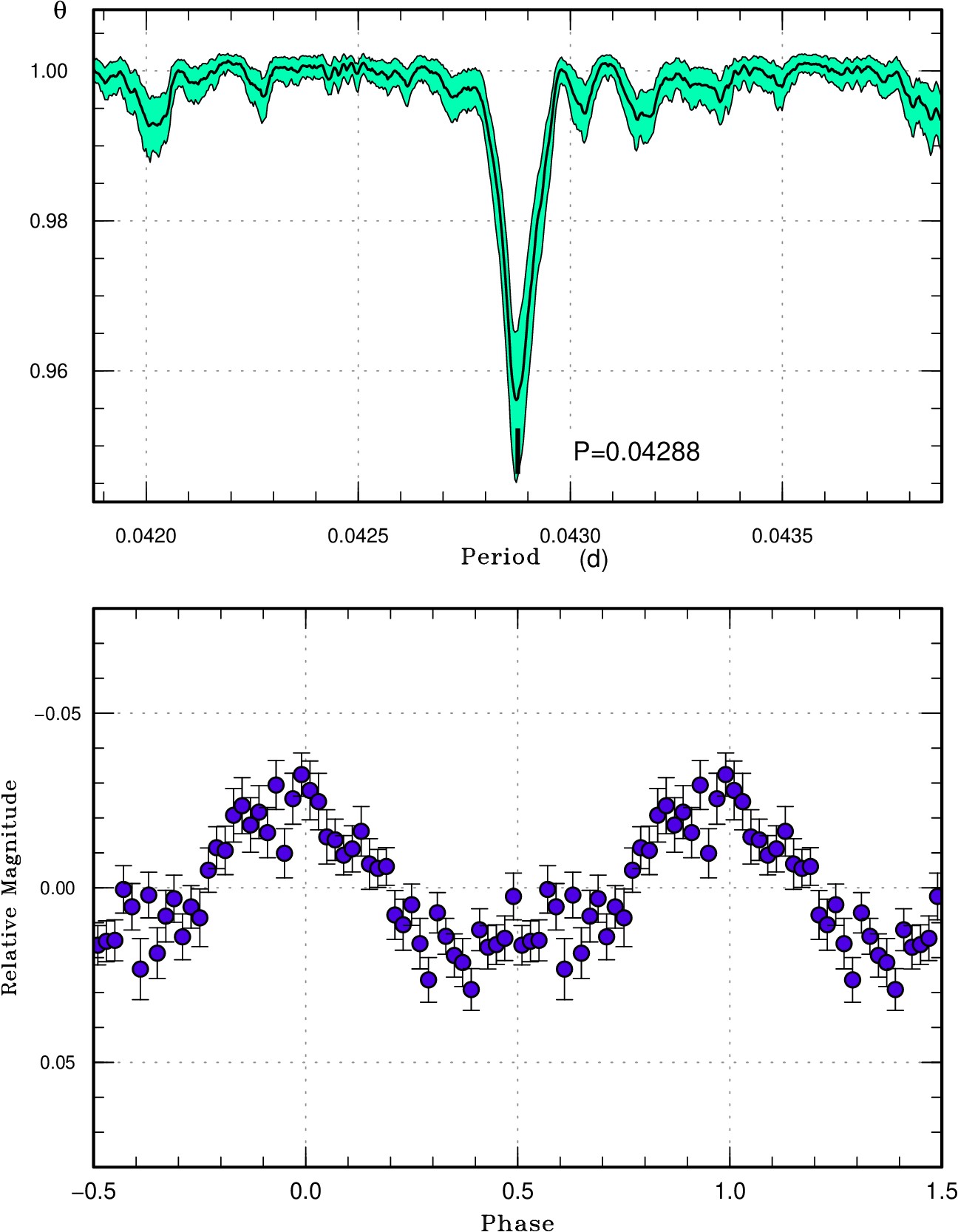}
\caption{
   Superhumps of ASASSN-22ak in 2023.
   (Upper): PDM analysis.  The bootstrap result using
   randomly contain 50\% of observations is shown as
   a form of 90\% confidence intervals in the resultant 
   $\theta$ statistics.
   (Lower): Phase plot.
}
\label{fig:shpdm2023}
\end{center}
\end{figure*}

\begin{figure*}
\begin{center}
\includegraphics[width=5.5cm]{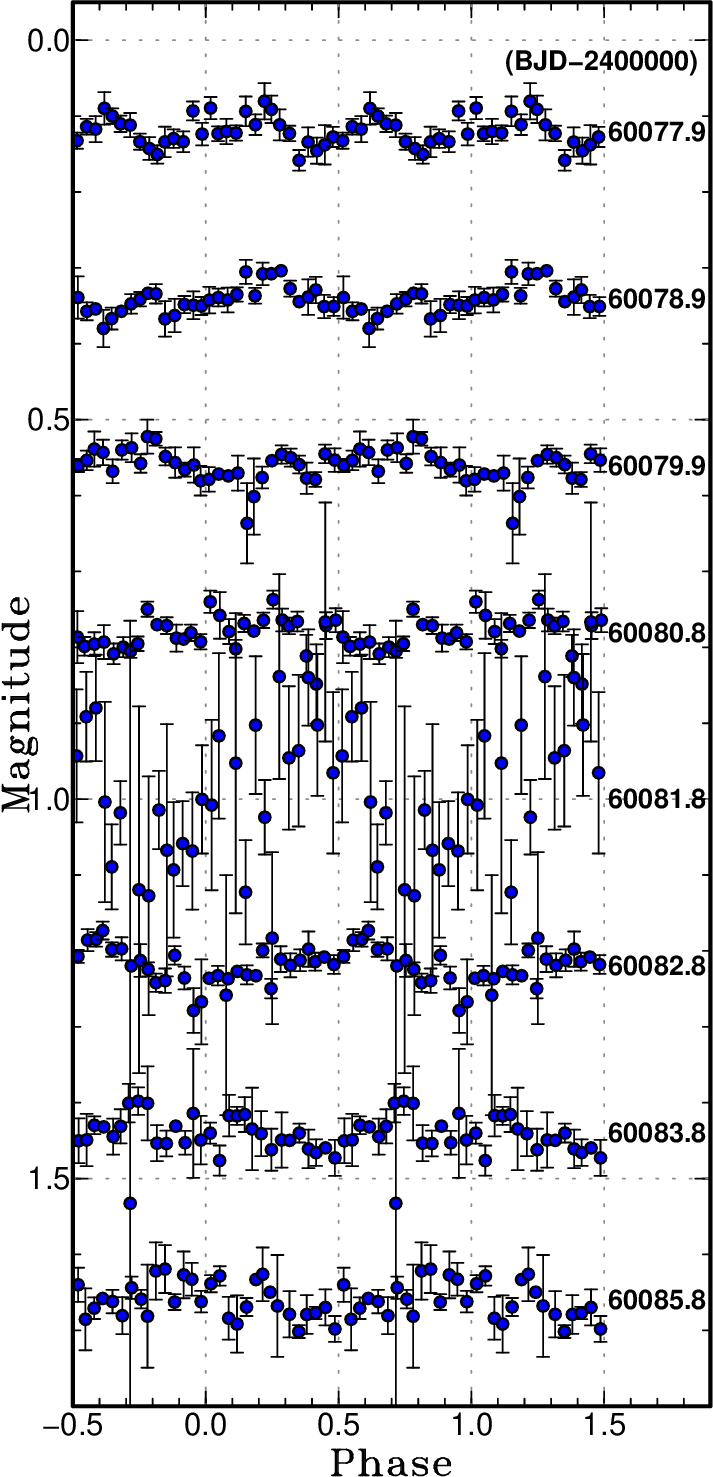}
\includegraphics[width=5.5cm]{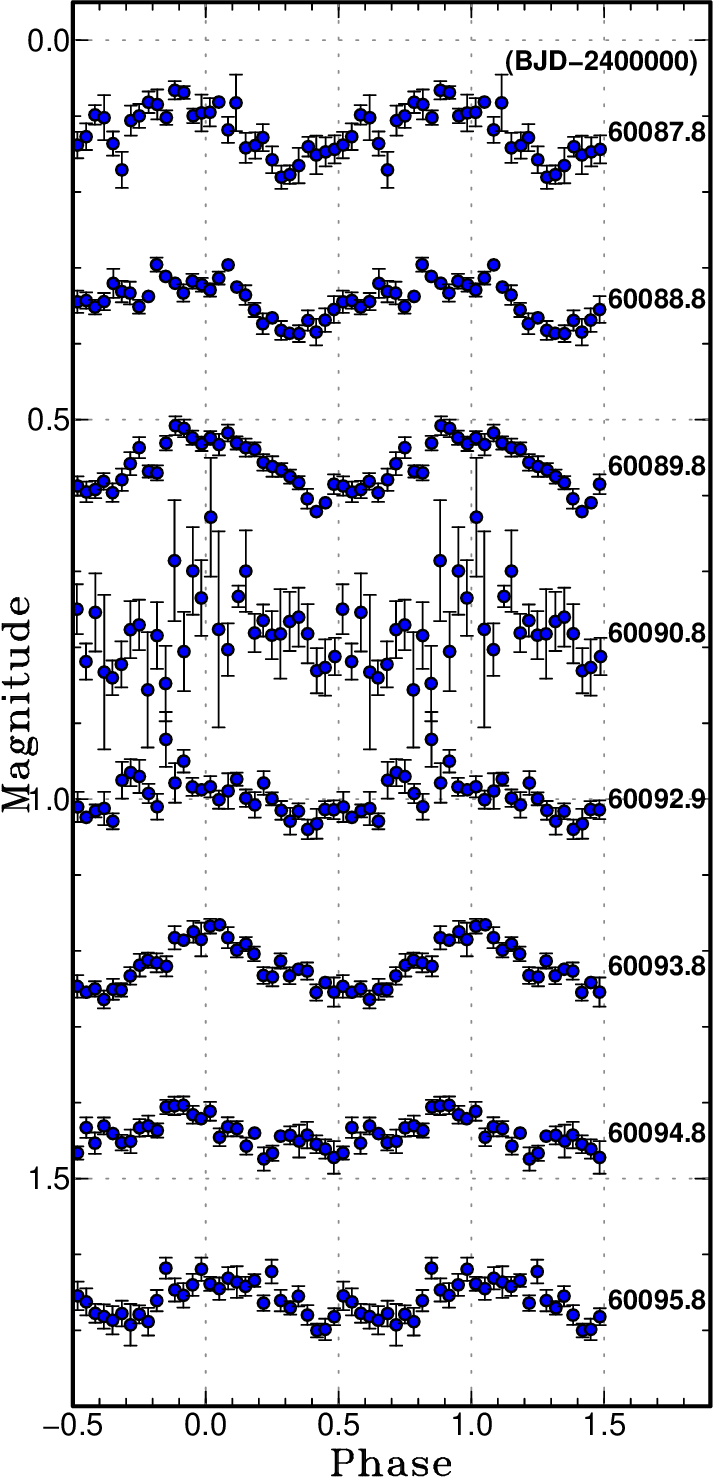}
\includegraphics[width=5.5cm]{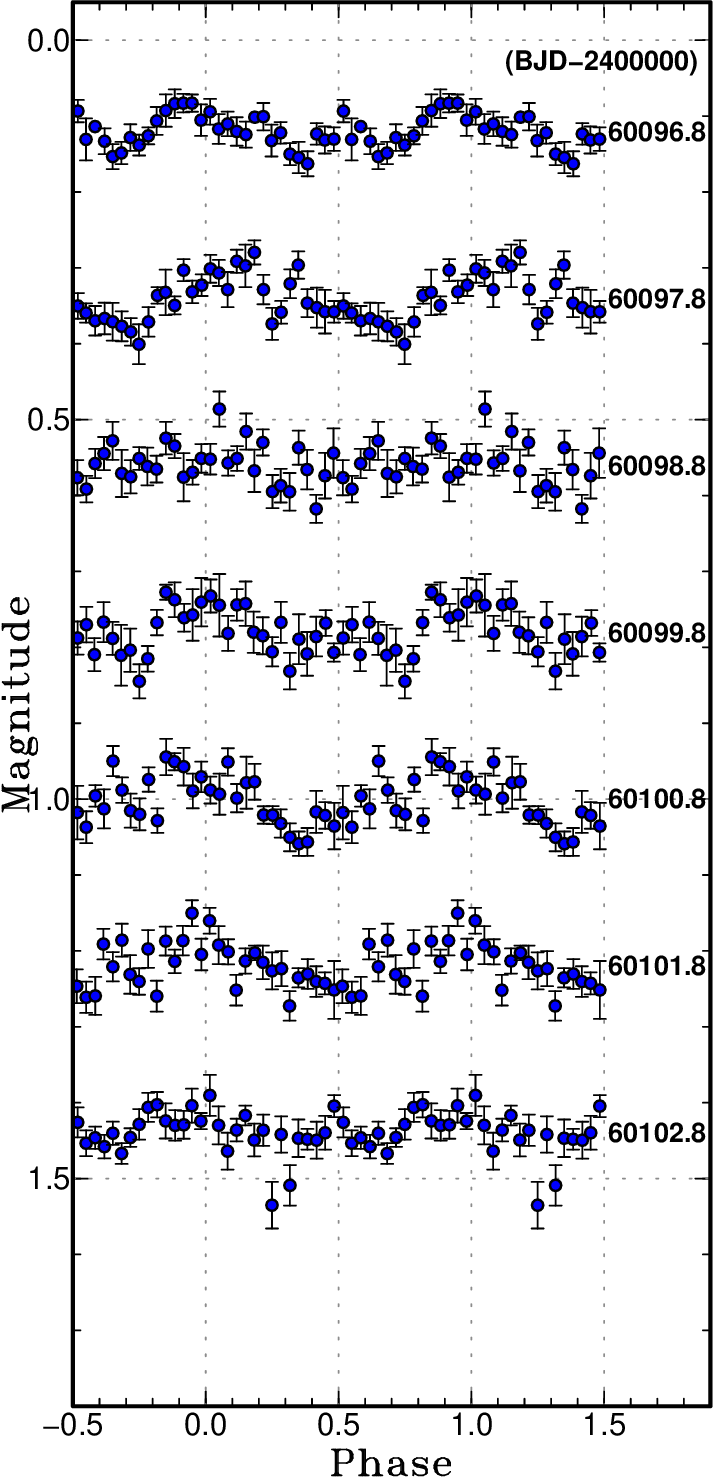}
\caption{
   Variation of the superhump profile in 2023.
   Superhumps of ASASSN-22ak in 2023.  The epoch is arbitrarily
   assumed to be BJD 2460089.800 and the period of 0.042876~d
   is used.
}
\label{fig:prof2023}
\end{center}
\end{figure*}

   We analyzed the best observed 2023 outburst.
We used locally-weighted polynomial regression (LOWESS: \cite{LOWESS})
to remove long-term trends.
The periods were determined using the phase dispersion
minimization (PDM: \cite{PDM}) method, whose errors were
estimated by the methods of \citet{fer89error,Pdot2}.
The result before the dip (2023 June 8, BJD 2460103), after
excluding the scattered data on 2023 May 17 (BJD 2460081--2460082)
is shown in figure \ref{fig:shpdm2023}.  The period obtained
by this analysis is 0.042876(3)~d.  The variation of the profiles
in 2023 is shown in figure \ref{fig:prof2023}.
The amplitude of the variations increased on BJD 2460088
(2023 June 23), which corresponded to temporary brightening
from the fading trend (see figure \ref{fig:lcout}).
Based on the amplitude variation correlated with the overall
trend similar to SU UMa stars \citep{Pdot} and the gradual
shift in the phase of peaks, we identified these variations
to be superhumps, not orbital variations.

\begin{figure*}
\begin{center}
\includegraphics[width=14cm]{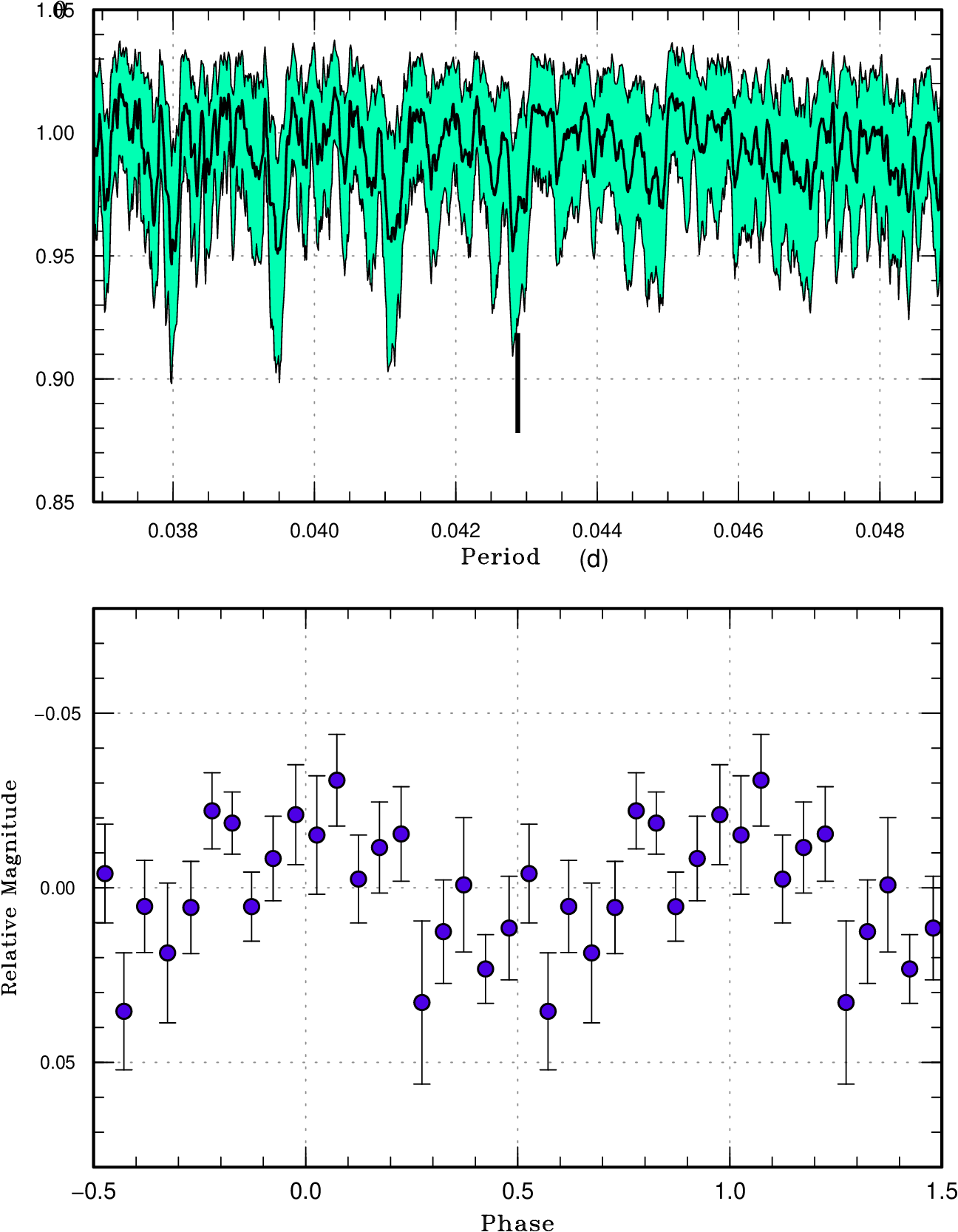}
\caption{
   Superhumps of ASASSN-22ak in 2022 December.
   (Upper): PDM analysis.  The tick is shown at the superhump
   period obtained by the 2023 observations.
   (Lower): Phase plot.
}
\label{fig:shpdm2022a}
\end{center}
\end{figure*}

   An analysis of less observed outburst in 2022 December
during the plateau phase is shown in figure \ref{fig:shpdm2022a}.
Note that these 7-night observations recorded only
the terminal portion of the outburst and the statistics
were not ideal.  The phase plot assumes a period of
0.042876~d, which is allowed as one of the aliases as
seen in the PDM analysis.

\section{Discussion}

\subsection{Comparison with hydrogen-rich WZ Sge stars}

   As we have seen, there was no evidence of an outburst
in ASASSN-22ak before 2022 (at least for seven years based on
ASAS-SN and Gaia observations).  The object suddenly became
active and repeated superoutbursts with cycle lengths of
132--188~d.  No very similar object has been known.  V3101 Cyg
is somewhat analogous in that it repeated four superoutbursts
(up to the time of the writing) following the 2019 large
outburst.  The case of V3101 Cyg is different in that short
rebrightenings were also observed \citep{tam20v3101cyg}.
The initial (2019) outburst of V3101 Cyg showed a relatively
rapidly fading phase, which is the viscous decay phase
characteristic to WZ Sge stars \citep{kat15wzsge}.
The initial (2022 January) outburst of ASASSN-22ak had
a similar feature, reaching $\sim$2~mag brighter than
subsequent outbursts and which apparently faded rapidly.
The second and third outbursts of ASASSN-22ak had similar,
but less distinct, features.  The same feature was almost
lacking in the fourth outburst (figure \ref{fig:lcout}).
These features suggest that the first outburst of ASASSN-22ak
was a strong WZ Sge-type one and that the second and third
ones were weaker WZ Sge-type ones,
although early superhumps \citep{kat15wzsge} were not
directly observed during any of these outbursts.

   The superhump period of 0.042876~d should be close to
the orbital period (see also discussions later).
This period is rather too short for a hydrogen-rich CV.
If ASASSN-22ak is a hydrogen-rich CV,
the orbital period should break the record of 0.0462583~d
in OV Boo \citep{lit07j1507,pat08j1507,uth11j1507,ohn19ovboo},
which is considered to be a population II CV.
We consider the possibility of ASASSN-22ak being
a population II CV less likely since the transverse velocity
of ASASSN-22ak is 20\% of OV Boo \citep{GaiaDR3} (but still with
a 28\% 1-$\sigma$ error in the Gaia parallax) and because
of the difference in the light curve (lack of short rebrightenings,
long durations of superoutbursts compared to supercycles)
from the hydrogen-rich V3101 Cyg.
ASASSN-22ak would then be more likely a hydrogen-depleted CV.
There are two possibilities.  It could be either an EI Psc
star (CV with an evolved core in the secondary but still with
considerable surface hydrogen) or an AM~CVn star in which
the surface hydrogen of the secondary is almost lost.
We consider these possibilities in more detail.

\subsection{Comparison with EI Psc stars in general}\label{sec:eipsc}

   EI Psc has an orbital period of 0.0445671(2)~d \citep{tho02j2329}
very similar to ASASSN-22ak.  EI Psc, however, has a hot, luminous
secondary \citep{tho02j2329}, whose quiescent color
(Gaia $GP-RP$=$+$0.88) is much redder than in ASASSN-22ak
($GP-RP$=$+$0.16).  Another EI Psc-type object V418 Ser
[superhump period 0.04467(1)~d] has $GP-RP$=$+$0.52 and this object
shows outbursts similar to hydrogen-rich CVs
\citep{Pdot7,vog21suumacycle}.
The properties of V418 Ser look different from those of ASASSN-22ak.
CRTS J174033.4$+$414756 (orbital period 0.045048~d) has
$GP-RP$=$+$0.43 and the outburst behavior
\citep{Pdot5,Pdot7,cho15j1740,ima18j1740}
appears moderately similar to ASASSN-22ak.
CRTS J174033.4$+$414756 indeed showed a bright WZ Sge-type outburst
in 2023 February after 5-yr quiescence (vsnet-alert 27373).\footnote{
   $<$http://ooruri.kusastro.kyoto-u.ac.jp/mailarchive/vsnet-alert/27373$>$.
}  Not sufficient time has passed since this outburst and
it is unknown whether CRTS J174033.4$+$414756 behaves like
ASASSN-22ak.  The known differences between CRTS J174033.4$+$414756
and ASASSN-22ak are that the former shows superhumps with
much larger amplitudes, which suggests a higher mass ratio
[$q$=0.077(5) was obtained by \citep{ima18j1740}], and
the redder color in quiescence.  Although CRTS J174033.4$+$414756
would be a good candidate for an already known object having
properties similar to ASASSN-22ak, particularly with a bright
superoutburst after 5-yr quiescence, the secondary in ASASSN-22ak
appears to be fainter and less massive.

\subsection{Comparison with CRTS J112253.3$-$111037}

\begin{figure*}
\begin{center}
\includegraphics[width=14cm]{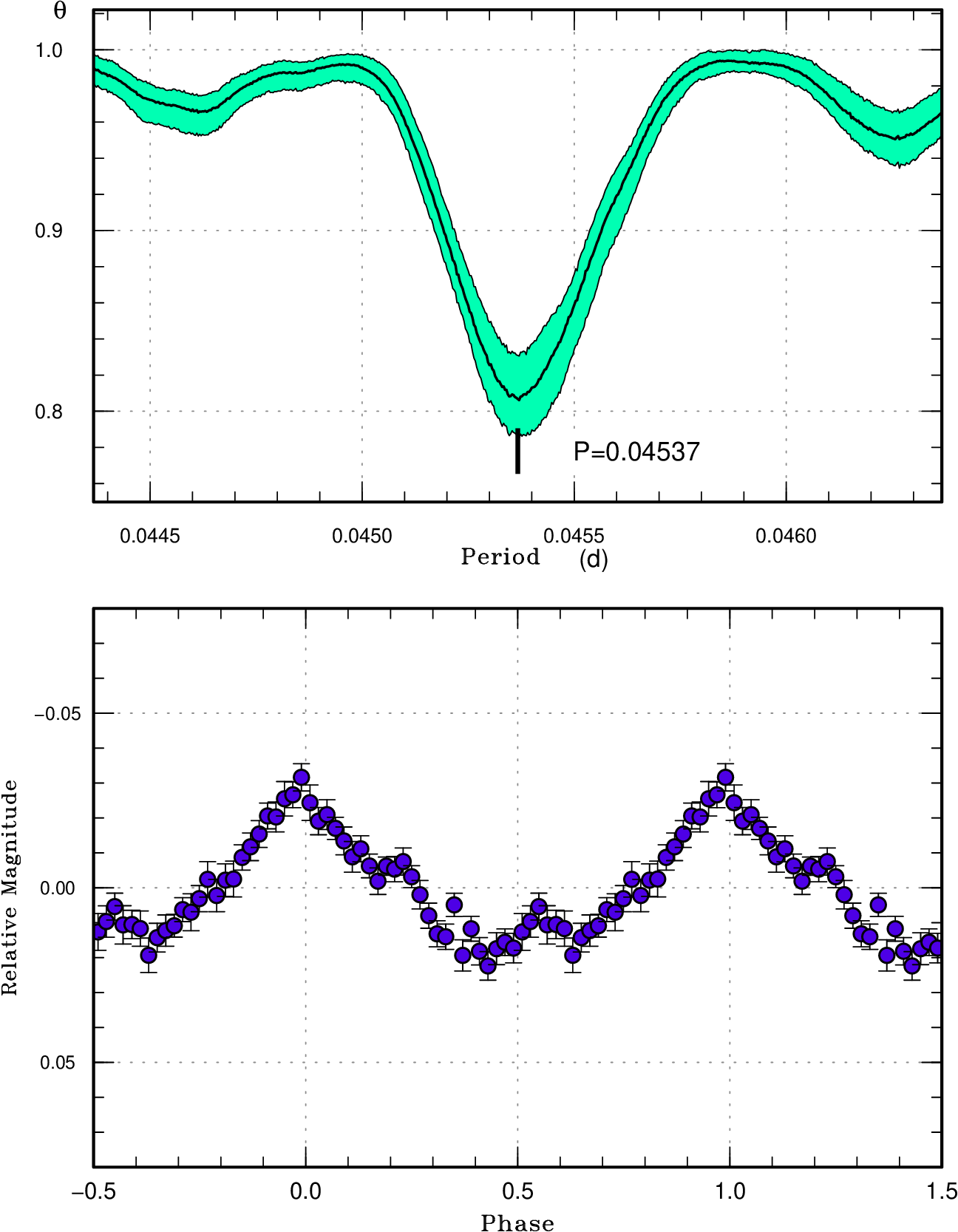}
\caption{
   Superhumps in CRTS J112253.3$-$111037 during the 2010 superoutburst.
   (Upper): PDM analysis.
   (Lower): Phase plot.
}
\label{fig:j1122shpdm2}
\end{center}
\end{figure*}

   The object most similar to ASASSN-22ak appears to be
CRTS J112253.3$-$111037 \citep{bre12j1122}.  This object
has an orbital period 0.04530~d and a very small fractional
superhump excess $\epsilon \equiv P_{\rm SH}/P_{\rm orb}-1$,
where $P_{\rm SH}$ and $P_{\rm orb}$ represent superhump and
orbital periods, respectively.  The secondary in
CRTS J112253.3$-$111037 was undetected in contrast to other
EI Psc stars.  The Gaia color $GP-RP$=$+$0.10 is also very
similar to that of ASASSN-22ak.
Although $P_{\rm SH}$ was reported in \citet{Pdot2}, this value
is vital to this discussion and we re-analyzed the data
in \citet{Pdot2}, in which the modern de-trending method was 
not yet employed.  The resultant period was 0.045409(9)~d
(figure \ref{fig:j1122shpdm2}).  This value corresponds to
$\epsilon$=0.0024(2).  In the treatment by \citet{bre12j1122},
old $\epsilon$-$q$ calibrations, which did not consider
the pressure effect, were used and they obtained
an exceptionally small $q$.  
Using the modern calibration in table 4 of \citet{kat22stageA}
considering the pressure effect (but calibrated using
hydrogen-rich systems), this $\epsilon$ corresponds to
$q$=0.043(1) assuming stage B superhumps [for superhump stages,
see \citet{Pdot}].
There remains a possibility that the observed superhumps
were stage C ones since observations only recorded the final
part of the outburst.  The periods of stage B superhumps
are generally longer by 0.5\% than those of stage C superhumps
in hydrogen-rich systems \citep{Pdot}.  If stage B superhumps
were missed and we only observed stage C superhumps, this
$q$ value would be an underestimate.  By artificially increasing
the superhump period by 0.5\%, the resultant $q$ becomes 0.058(1),
which should be regarded as the upper limit.
In actual WZ Sge stars, stage C tends to be missing
\citep{Pdot,kat15wzsge}, and we consider that the first value
[$q$=0.043(1)] is expected to be closer to the real one.

   CRTS J112253.3$-$111037 is also similar to ASASSN-22ak
in terms of the low frequency of outbursts \citep{bre12j1122}.
There was no information how the 2010 outburst in
CRTS J112253.3$-$111037 started due to an $\sim$50~d
observational gap in the CRTS data \citep{CRTS} and
it is unknown whether CRTS J112253.3$-$111037 showed
a sharp peak or a viscous decay phase.
No repeated superoutbursts like ASASSN-22ak, however, appear to
have been present since then.
It might be interesting to note that ATLAS and ASAS-SN data
show that CRTS J112253.3$-$111037 showed brightening with
a broad peak reaching $g$=17.8 around 2022 June 6 (BJD 2459737).
The entire event lasted $\sim$15~d and this may be similar
to the enhanced quiescent activity in the AM CVn star
NSV 1440 \citep{kat23nsv1440}, possibly signifying
the similarity to AM CVn stars.

   The small amplitude of superhumps (0.05~mag) in
CRTS J112253.3$-$111037 is also similar to ASASSN-22ak (0.05~mag),
implying a similarly low $q$ in ASASSN-22ak.

   \citet{bre12j1122} suggested a possibility that
CRTS J112253.3$-$111037 had already evolved past its period
minimum based on \citet{pod03amcvn} and that its secondary
can be semidegenerate.  Although this conclusion was apparently
partly based on $q$ smaller than the one obtained in the present paper,
we agree that both ASASSN-22ak and CRTS J112253.3$-$111037
are evolving close to AM CVn stars since the properties of
these objects are very different from other EI Psc objects
with similar orbital periods (subsection \ref{sec:eipsc}).
ASASSN-22ak may have already lost hydrogen and it may even be
an AM CVn star.  If this is the case, ASASSN-22ak breaks
the longest record of orbital periods in AM CVn stars showing
a genuine superoutburst [see also the discussion in
\citet{kat23nsv1440}; superhump period of 0.0404--0.0415~d
in ASASSN-21au = ZTF20acyxwzf
(vsnet-alert 25369;\footnote{
   $<$http://ooruri.kusastro.kyoto-u.ac.jp/mailarchive/vsnet-alert/25369$>$.
} \cite{iso21asassn21auatel14390,riv22asassn21au})].
Another AM CVn star with a long orbital period
[PNV J06245297$+$0208207 in 2023 \citep{mae23j0624atel15849}:
superhump period 0.035185(8)~d
(vsnet-alert 27353\footnote{
   $<$http://ooruri.kusastro.kyoto-u.ac.jp/mailarchive/vsnet-alert/27353$>$.
})]
showed a superoutburst very similar to ASASSN-21au.
This morphology of superoutbursts appears to be common
to AM~CVn stars with long orbital periods and the possibility
of ASASSN-22ak as being an AM CVn star might be less likely.
We leave this question open since the outburst properties
were so unusual in ASASSN-22ak.
In any case, spectroscopy of ASASSN-22ak to determine
the hydrogen and helium content and to determine
the orbital period is very much desirable.
The addition of ASASSN-22ak seems to strengthen the idea
that cataclysmic variables could be the dominant
progenitors of AM CVn binaries
\citep[see, e.g.,][]{sak23amcvnevol,bel23amcvnevol}.

   The EI Psc-type objects treated in this paper for comparisons
with ASASSN-22ak are summarized in table \ref{tab:objcomp}.
The mean superhump amplitude for EI Psc was obtained from
the data in \citet{uem02j2329letter}.
The superhump amplitudes for V418 Ser and CRTS J174033.4$+$414756
were from \citet{Pdot7} and \citet{ima18j1740}, respectively.
Although CRTS J174033.4$+$414756 showed the initial phase of large
superhump amplitudes \citep{ima18j1740}, no such a phase was
recorded in ASASSN-22ak.  The $q$ values from $\epsilon$ assuming stage B
were obtained by the method in \citet{kat22stageA}.

\begin{table*}
\caption{Comparison of objects treated in this paper.}\label{tab:objcomp}
\begin{center}
\begin{tabular}{cccccc}
\hline
Object & Period (d)\commenta & $BP-RP$ & $M_G$ & $q$\commentb & Superhump amplitude (mag) \\
\hline
EI Psc & 0.0445671 & $+$0.88 & 10.0 & 0.171 & 0.16 \\
V418 Ser & 0.04467(1)$^\textrm{s}$ & $+$0.52 & 10.3(5) & -- & 0.06 \\
CRTS J174033.4$+$414756 & 0.045048 & $+$0.43 & 11.0(3) & 0.077(5) & 0.17--0.03 \\
CRTS J112253.3$-$111037 & 0.04530(1) & $+$0.10 & 11.1(10) & 0.043(1) & 0.05 \\
ASASSN-22ak & 0.042876(3)$^\textrm{s}$ & $+$0.16 & 12.1(5) & -- & 0.05 \\
\hline
  \multicolumn{6}{l}{\commenta Superscript s refers to superhump period.} \\
  \multicolumn{6}{l}{\commentb Value estimated from $\epsilon$ assuming stage B.} \\
\end{tabular}
\end{center}
\end{table*}

\subsection{Pre-outburst dormancy and repeated superoutbursts}

   Repeated long superoutbursts with short recurrence times
is the unique feature of ASASSN-22ak.  In the case of
(hydrogen-rich) V3101 Cyg, some of post-superoutburst
rebrightenings may have been caused by the matter in the disk
left after the main superoutburst \citep{tam20v3101cyg}.
Repeated superoutbursts appear to be more easily explained
if the mass-transfer rate increased after the initial outburst
\citep{ham21DNrebv3101cyg}.  This increase in the mass transfer
may either have been caused by irradiation of the secondary
by the initial outburst \citep{ham21DNrebv3101cyg},
or it could have been that the quiescent viscosity of the disk
before the initial outburst was simply extremely low to accumulate
a large amount of mass in the disk and that the mass-transfer
rate and the quiescent viscosity is simply returning to
the normal value of this object after the initial outburst.

   In the case of ASASSN-22ak, the initial outburst was
not as strong as in V3101 Cyg, although the peak was bright,
and the mechanism may be different from the case of
V3101 Cyg.  In ASASSN-22ak, $q$ would be smaller than in
V3101 Cyg (as inferred from the smaller amplitude of
superhumps and from the analogy with CRTS J112253.3$-$111037)
and the weaker tidal effect would make it more difficult to
maintain superoutbursts in contrast to V3101 Cyg.
Although there have been a suggestion that smaller $q$
can lead to premature quenching of superoutbursts
\citep[see e.g.,][]{hel01eruma}, there is no established
theory when superoutbursts end.  Although this premature
quenching of superoutbursts might explain the repeated
superoutbursts with relatively short intervals, the lack of
post-superoutburst rebrightenings in ASASSN-22ak might be
problematic.  It may be that the hydrogen
depletion in the disk of ASASSN-22ak is not as strong as
AM CVn stars and long superoutbursts are easier to
maintain than in almost pure helium disks.  A combination
of effects of all these circumstances, unusual for
ordinary CVs, should be a challenging target for
theorists working with the disk-instability model.

   The pre-outburst dormancy might be easier
to explain in ASASSN-22ak.  In contrast to V3101~Cyg, 
which is expected to have a fully convective secondary,
ASASSN-22ak has an evolved core and a magnetic dynamo can still work 
\citep[see e.g.,][]{sak23amcvnevol} and is probably necessary
to form the observed AM CVn stars within reasonable time.
With such a dynamo, the instantaneous mass-transfer rate can
be different from the secular average, as seen in the spread of
absolute magnitudes in CVs above the period gap \citep{deb18CVgaia}
and the presence of VY~Scl stars.
There is also a possibility that the quiescent viscosity
of the disk before the initial outburst was simply very low and
the viscosity increased after the outburst as proposed by
\citet{osa01egcnc,mey15suumareb} for hydrogen-rich WZ Sge stars.
This explanation, however, might face a difficulty
to realize a very quiet, low-viscosity disk when the secondary
has a seed magnetic field, which may increase
the quiescent viscosity of the disk via the magneto-rotational
instability (cf. \cite{mey99diskviscosity};
but see also \cite{ish01rzleo}).
High and low states in polars (AM Her stars: \cite{cro90polarreview})
may provide additional insight.
EF Eri has a brown-dwarf secondary
\citep[][and the references therein]{sch10eferi} and a strong
magnetic activity cycle as in CVs above the period gap
is not expected.  This object showed (and still showing)
a long-lasting high state (just like ``awakening'')
starting from 2022 December (vsnet-alert 27205).\footnote{
   $<$http://ooruri.kusastro.kyoto-u.ac.jp/mailarchive/vsnet-alert/27205$>$.
} Since polars do not have an accretion disk, storage of mass
in the disk before the active (high) state, as in WZ~Sge stars,
is impossible.  There could be a reservoir of additional angular
momentum other than the disk, and this might also explain
the dormany/waking-up phenomena in dwarf novae.

\section*{Acknowledgements}

This work was supported by JSPS KAKENHI Grant Number 21K03616.
The authors are grateful to the ASAS-SN, ATLAS and Gaia teams
for making their data available to the public.
We are also grateful to Naoto Kojiguchi for helping downloading
the ZTF and Gaia data and Yusuke Tampo, Junpei Ito and
Katsuki Muraoka for converting the data reported to
the VSNET Collaboration.

This work has made use of data from the Asteroid Terrestrial-impact
Last Alert System (ATLAS) project.
The ATLAS project is primarily funded to search for
near earth asteroids through NASA grants NN12AR55G, 80NSSC18K0284,
and 80NSSC18K1575; byproducts of the NEO search include images and
catalogs from the survey area. This work was partially funded by
Kepler/K2 grant J1944/80NSSC19K0112 and HST GO-15889, and STFC
grants ST/T000198/1 and ST/S006109/1. The ATLAS science products
have been made possible through the contributions of the University
of Hawaii Institute for Astronomy, the Queen's University Belfast, 
the Space Telescope Science Institute, the South African Astronomical
Observatory, and The Millennium Institute of Astrophysics (MAS), Chile.

We acknowledge ESA Gaia, DPAC and the Photometric Science Alerts Team
(http://gsaweb.ast.cam.ac.uk/\\alerts).

\section*{List of objects in this paper}
\xxinput{objlist.inc}

\section*{References}

We provide two forms of the references section (for ADS
and as published) so that the references can be easily
incorporated into ADS.

\newcommand{\noop}[1]{}\newcommand{\hyphalt}{-}

\renewcommand\refname{\textbf{References (for ADS)}}

\xxinput{asn22akaph.bbl}

\renewcommand\refname{\textbf{References (as published)}}

\xxinput{asn22ak.bbl.vsolj}

\end{document}